\documentclass[aps,prl,groupedaddress,showpacs,floatfix,nofootinbib,preprintnumbers,notitlepage,twocolumn]{revtex4-2}

\usepackage[T1]{fontenc}
\usepackage{lmodern}
\usepackage{amsmath,amssymb,bm,mathtools,mathrsfs}
\usepackage[letterpaper,top=1.75cm,bottom=1.75cm,left=1.65cm,right=1.65cm,columnsep=0.55cm]{geometry}
\usepackage{microtype}
\usepackage{natbib}
\usepackage{xcolor}
\usepackage[colorlinks=true,citecolor=blue!55!black,linkcolor=blue!55!black,urlcolor=blue!55!black]{hyperref}
\hypersetup{
  pdftitle={From Nuclear Many-Body Correlations to Energy Detector 
Correlators},
  pdfauthor={João Barata}
}

\newcommand{\cE}{\mathcal{E}}
\newcommand{\avg}[1]{\left\langle #1\right\rangle}

\newcommand{\nn}{\nonumber \\}

\begin{document}

\title{From Nuclear Many-Body Correlations to Energy Detector Correlators}

\author{Jo\~{a}o Barata}
\affiliation{Theoretical Physics Department, CERN, 1211 Geneva 23, Switzerland}

\author{Giuliano Giacalone}
\affiliation{Theoretical Physics Department, CERN, 1211 Geneva 23, Switzerland}

\preprint{CERN-TH-2026-203}

\begin{abstract}
Relativistic nuclear collisions have opened an experimental arena for studying many-body correlations in nuclear ground states. However, the connection between initial-state correlations and final-state multi-particle observables measured at colliders is not yet
formulated as a systematically improvable matching problem. We show that detector correlators built from asymptotic energy flows provide a natural framework for realizing such a construction. 
In particular, we express the asymptotic energy flow as a functional of the early-time stress tensor, and expand it in suitable modes to recover the familiar linear hydrodynamic relations at leading order. Then, motivated by small-$x$ QCD, we map angular projections of detector correlators to multipole-operator correlators computed in the colliding nuclei. We thus establish a systematic formalism for matching long-wavelength correlations between incoming and outgoing QCD states 
in high-energy hadronic collisions.
\end{abstract}

\maketitle

One of the most far-reaching and least anticipated legacies of the high-energy nuclear collision programs at the Relativistic Heavy Ion Collider (RHIC) and the Large Hadron Collider (LHC) is the realization that these experiments act as microscopes of the structure of the colliding nuclear ground states \cite{Jia:2022ozr}. The first indications came with the discovery of elliptic-flow fluctuations \cite{PHOBOS:2005gex,PHOBOS:2006dbo}, originating from the finite, mesoscopic nature of the colliding nuclei \cite{Miller:2003kd,Bhalerao:2006tp,PHOBOS:2007vdf}. The picture became firmly established following the recognition that the same geometric argument also explains higher-order harmonics of the measured particle distributions, such as triangular flow \cite{Alver:2010gr,Teaney:2010vd}. This paradigm thus holds that collective expansion converts spatial anisotropy in the produced medium into momentum-space correlations among emitted hadrons \cite{Heinz:2013th,Ollitrault:2023wjk}.

Later, beginning with measurements of U+U collisions at RHIC \cite{STAR:2015mki,STAR:2024wgy,STAR:2025elk}, and subsequently extending to Xe+Xe collisions at the LHC \cite{ALICE:2018lao,CMS:2019cyz,ATLAS:2019dct,ATLAS:2022dov,ALICE:2024nqd,CMS:2025opi}, isobar collisions \cite{STAR:2021mii}, and recent light-ion collisions \cite{ATLAS:2025nnt,ALICE:2025luc,CMS:2025tga,ATLAS:2026zgq,LHCb:2025ixz}, it has become clear that flow observables are sensitive not only to the finite number of nucleons but also to spatial correlations among them. This sensitivity manifests itself through nontrivial departures from a smooth system-size dependence of observables, and has been successfully interpreted through semi-classical descriptions of nuclei \cite{Jia:2021tzt,Jia:2021qyu}, whereby deformed or clustered intrinsic configurations are assigned to effectively model the influence of collective inter-nucleon correlations \cite{Giacalone:2017dud,Zhang:2021kxj,Nijs:2021kvn,Zhao:2022uhl,Ryssens:2023fkv,YuanyuanWang:2024sgp,Zhao:2024feh,Giacalone:2024luz,Giacalone:2024ixe,Fortier:2024yxs,Mantysaari:2024uwn,Liu:2025zsi,Li:2025hae,Xu:2025cgx,Zhang:2025zrm,Mehrabpour:2026lhj,MenonKavumpadikkalRadhakrishnan:2026tfn}. High-energy collisions emerge, thus, as tools for imaging the low-energy structure of the collided isotopes.

This opportunity motivates the identification of nuclear ground-state properties probed by multiparticle correlations within a quantum many-body framework \cite{Giacalone:2023hwk}. This requires converting notions such as \textit{deformation} to expectation values of operators acting on the nuclear states. Following~\cite{Duguet:2025hwi}, consider collisions in which all nucleons participate, and denote by
$V_n\propto\int d\phi\,(dN/d\phi)e^{in\phi}$ the complex Fourier coefficient of the final hadron distribution. For $n\geq2$, the measured mean-squared flow harmonic can be related to a ground-state multipole-operator correlator,
\begin{align}
\label{eq:introOperator}
\left\langle V_n \, V_n^\ast   \right\rangle_{\rm final\,state}
\propto
\left\langle \widehat {\mathcal{Q}}_n  \widehat {\mathcal{Q}}_n^\dagger 
\right\rangle_{\rm nucleus} +~~\ldots,
\end{align}
where $\widehat {\mathcal{Q}}_{n} = \sum_i^A r_i^n e^{in\phi_i}$ is the maximally projected multipole operator for a system of $A$ nucleons, the ellipses imply corrections to higher-body and higher-radial operators, while $(r,\phi)$ denotes coordinates relative to the nuclear center of mass in the plane transverse to the beam direction. Equation~\eqref{eq:introOperator} thus places collider observables alongside matrix elements familiar from low-energy nuclear physics \cite{Kumar:1972zza,Cline:1986ik,Poves:2019byh,Henderson:2020yql,Garrett:2021kfb}, opening a research program for analyzing nuclear many-body correlations at colliders \cite{Duguet:2025hwi,Blaizot:2025scr,Ke:2025tyv,Blaizot:2025bfu,Bofos:2026huw,Mehrabpour:2026yuc,Bofos:2026nmg,Xu:2026llz,Giacalone:2026fat,Ding:2026foi}.

However, these advances expose limitations that should be overcome to fully establish high-energy collisions as discovery tools for nuclear many-body dynamics.

First, the derivation of Eq.~\eqref{eq:introOperator} takes as input the hydrodynamic response relation $V_n=\kappa_n\boldsymbol{\varepsilon}_n+\ldots$,
where $\kappa_n$ encodes the collision dynamics and $\boldsymbol{\varepsilon}_n$ is the $n$th-order eccentricity of the initial energy density, with the ellipsis containing higher-radial moments and nonlinear mode couplings \cite{Teaney:2010vd,Teaney:2012ke,Sousa:2024msh}. Therefore, imposing the relation \textit{assumes without deriving} the form of $\boldsymbol{\varepsilon}_n$ as the leading variable and that of higher-order corrections, which should instead emerge in a controlled initial-to-final matching formalism.

Second, the endpoints of Eq.~\eqref{eq:introOperator} are formulated in different languages. The incoming nucleus is described through a nucleon-level stress tensor density, whereas $V_n$ is conventionally constructed from the number-weighted distribution of asymptotic hadrons. The inclusive hadron number is not the flux of a locally conserved QCD current, and the fully differential hadron distribution is not determined by $T^{\mu\nu}$ alone. Its calculation requires additional microscopic information about, e.g.,  particlization and resonance decays. These effects can be incorporated into $\kappa_n$, which then becomes a response coefficient for a chosen hadronic observable rather than, by itself, an operator-level matching coefficient connecting the incoming and outgoing states.

\begin{figure*}[t]
    \centering
    \includegraphics[width=\linewidth]{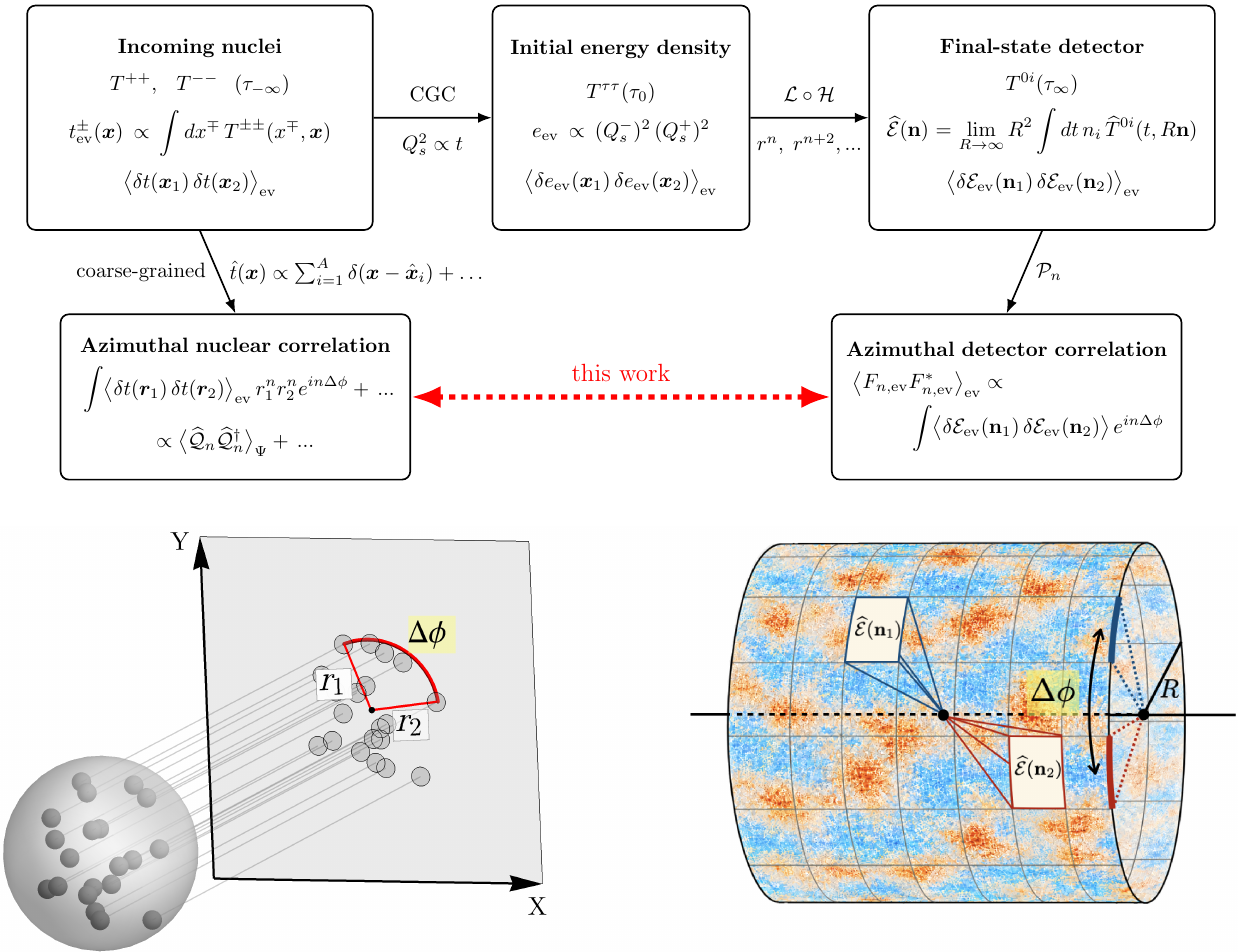}
    \caption{Summary of the end-to-end chain developed in this work. The incoming nuclear stress tensors define transverse thickness fields, $t_{\rm ev}^{\pm}(\bm x)$, and their event-by-event correlations. Upon coarse-graining the nuclear degrees of freedom, the azimuthal projection of the corresponding two-point function is expressed in terms of nuclear multipole operators, with the leading contribution proportional to $\langle\widehat{\mathcal Q}_n\widehat{\mathcal Q}_n^\dagger\rangle_\Psi = \bigl \langle \sum_{ij}r_i^n r_j^n e^{in\Delta\phi} \bigr \rangle_\Psi$. The collision prescription for the event energy density profile at $\tau_0$ is then represented by the leading dense-dense CGC scaling, $e \propto (Q_s^-)^2(Q_s^+)^2$ with $Q_s^2\propto t$. The subsequent evolution through $\mathcal H$ is organized in the long-wavelength expansion retaining the lowest radial weights ($r^n,\ldots$), followed by the asymptotic light-ray map $\mathcal L$ that determines the event-wise energy distribution measured by a distant detector. Its $n$th angular projection gives the complex harmonic $F_{n,\rm ev}$, so that $\langle F_{n,\rm ev}F_{n,\rm ev}^{*}\rangle_{\rm ev}$ is the $n$th Fourier component of the energy-energy correlator. The lower panels illustrate the relative azimuthal angle, $\Delta\phi$, on the nuclear and detector sides, which is the focus of our discussion.}
    \label{figure}
\end{figure*}

In this Letter, we address both gaps. We argue that the natural theoretical construction emerges by introducing observables that admit the relevant operator language. These are asymptotic energy flow \textit{detector} operators~\cite{Hofman:2008ar,Basham:1978zq,Sveshnikov:1995vi,Korchemsky:1999kt},
\begin{equation}
 \widehat{\mathcal{E}}(\mathbf{n})
 =\lim_{R\to\infty}R^2\int_{0}^{\infty}d t\,
 n_i\widehat T^{0i}(t,R\,\mathbf{n}),
 \label{eq:energyflow}
\end{equation}
for a detector oriented along the unit vector $\mathbf{n}$ 
at a radial distance $R$. Such operators correspond to idealized theoretical calorimeters, linking collider experiments to QCD theory, see~\cite{Moult:2025nhu} for a recent review. Their connection with the infrared properties of the bulk matter produced in heavy-ion collisions was first alluded to in~\cite{Krasnitz:1998ns}, and substantially developed in recent works~\cite{Barata:2024ukm,Barata:2025fzd,Bossi:2024qho,Barata:2026pgh,Yang:2023dwc,Zhao:2025ogc,Duan:2026icj,Yiyang27}. In~\cite{Yiyang27}, in particular, event-wide EECs are introduced as new probes of the hydrodynamic medium. Here, we add a new link to the structure of the incoming nuclei, completing the end-to-end chain as summarized in Fig.~\ref{figure}.

Detector correlation functions,  $\langle  \widehat {\cE}(\mathbf{n}_1)\cdots \widehat \cE(\mathbf{n}_k)\rangle$, are evaluated on the final state of hadronic scattering events. For $k=2$ we introduce the energy-energy correlator (EEC)
\begin{equation}
 \mathcal G_2\equiv
 \avg{\widehat\cE(\mathbf{n}_1)\widehat\cE(\mathbf{n}_2)}
 -\avg{\widehat\cE(\mathbf{n}_1)}\avg{\widehat\cE(\mathbf{n}_2)}.
 \label{eq:EEC}
\end{equation}
As commonly done in the analysis of high-multiplicity events, one splits the average in two parts. First we average over an individual event, that is, consider the average energy hitting an asymptotic detector in a single collision event. Second, we take the statistical average over events. Introducing $\cE_{\mathrm{ev}}(\mathbf{n})\equiv \avg{\widehat{\cE}(\mathbf{n})}_{\mathrm{fixed\,ev}}$, and $\delta\cE_{\mathrm{ev}} =\cE_{\mathrm{ev}}-\avg{\cE_{\mathrm{ev}}}_{\mathrm{ev}}$, the EEC becomes the sum of covariances
\begin{equation}
 \mathcal G_2= \mathcal G_2^{\mathrm{geom}}  + \avg{\mathcal G_{2}^{\mathrm{fixed}}}_{\mathrm{ev}} \, ,
 \label{eq:EECdecomposition}
\end{equation}
where $\mathcal{G}_2^{\rm geom} = \avg{
 \delta\cE_{\mathrm{ev}}(\mathbf{n}_1)
 \delta\cE_{\mathrm{ev}}(\mathbf{n}_2)}_{\mathrm{ev}}$, on which we focus in this paper, and analogously for $\avg{ \mathcal G_{2}^{\mathrm{fixed}}}_\mathrm{ev}$.

 In high-multiplicity hadronic collisions, within the soft sector of produced hadrons and at angular separations larger than microscopic 
 scales, $\mathcal{G}_2^{\rm geom}$ is the \textit{classical} contribution that measures correlations associated with the collective or hydrodynamic flow~\cite{Barata:2026pgh,Cuomo:2025pjp,Chicherin:2023gxt}. Within the hydrodynamics flow paradigm, it corresponds to the long-range part of the two-particle correlations 
 as a function of (relative) detector angles \cite{Luzum:2011mm}. The second term, $\mathcal G_{2}^{\mathrm{fixed}}$, contains instead thermal, hadronization, and other fluctuations at fixed event geometry, which mainly contribute at smaller angular separations, sensitive to the microscopic details of the bulk matter~\cite{Barata:2026pgh}. In the language of the flow analyses, this contributes to the so-called \textit{non-flow}, usually removed through subtraction procedures to isolate the geometric long-range component \cite{Feng:2024eos}. The separation of these two contributions is highly non-trivial, even in theories where there is a greater theoretical control~\cite{Cuomo:2025pjp,Chicherin:2023gxt} compared to QCD, and requires a detailed understanding of the departure from equilibrium~\cite{Delacretaz:2018cfk,Barata:2026pgh}. Here we do not tackle the issue of the separation between flow and non-flow correlations, and consider a generic hydrodynamic scenario where the contribution $\avg{ \mathcal G_{2}^{\mathrm{fixed}}}_\mathrm{ev}$ to Eq.~(\ref{eq:EECdecomposition}) is negligible. For an explicit quantitative example of such contributions across collisional systems see~\cite{Yiyang27}.

We can now connect this discussion to observables akin to those measured in anisotropic-flow analyses. Defining detector orientations through pseudorapidity, $\eta=-\ln\tan(\theta/2)$, and azimuthal angle, $\phi$, the event energy flow becomes~\cite{Barata:2026pgh}:
\begin{equation} \cE_{\mathrm{ev}}(\eta,\phi) = \cosh^3\eta\, \frac{dE_T}{d\phi d\eta}\biggl|_{\rm ev}, 
\end{equation} 
where $E_T \equiv E\sin\theta = E/\cosh\eta$. To simplify the notation, we work at midrapidity, $\eta=0$, and we denote $\mathcal{F}_{\rm ev}(\phi) = (\frac{dE_T}{d\phi d\eta}\bigl|_{\rm ev})_{\eta=0}$. Introducing dimensionful Fourier coefficients
$F_{n,\mathrm{ev}}\equiv
\int_0^{2\pi}d\phi\,
e^{in\phi}\mathcal F_{\mathrm{ev}}(\phi)$,
we write
\begin{equation}
 \mathcal F_{\mathrm{ev}}(\phi)
 =
 \frac{1}{2\pi}
 \sum_{n\in\mathbb Z}
 F_{n,\mathrm{ev}}e^{-in\phi}.
 \label{eq:energyFlowExpansion}
\end{equation}
The zeroth coefficient,
$F_{0,\mathrm{ev}}
=\left.dE_T/d\eta\right|_{\eta=0,\mathrm{ev}}$,
is the transverse energy per unit pseudorapidity at midrapidity. Rotational invariance of the event ensemble implies $\langle F_{n,\mathrm{ev}}\rangle_{\mathrm{ev}}=0$ for $n>0$, and diagonalizes the covariance of different harmonics. Therefore, at midrapidity we can write 
\begin{align} 
\mathcal G_2^{\mathrm{geom}} &=\frac{\left \langle (\delta F_{0, \rm ev})^2 \right \rangle_{\mathrm{ev}}}{(2\pi)^2}  + \sum_{n \in \mathbb{Z} \setminus \{0\}} \frac{\left\langle f_n^2 \right\rangle_{\mathrm{ev}} }{(2\pi)^2}\, e^{-i n(\phi_1-\phi_2)} \, , 
\label{eq:EECharmonics} 
\end{align} 
where $f_n = |F_{n, \rm ev}|$. We see that $\langle f_n^2 \rangle$ is the dimensionful, unnormalized energy-flow counterpart of the traditional mean-squared flow harmonic, $\langle v_n^2 \rangle \equiv v_n\{2\}^2$. Generalizations  within a finite pseudorapidity window are given in the Supplement. Equation \eqref{eq:EECharmonics} is the first result of this Letter. It gives the angular projection of the stress tensor correlator evaluated in the geometric regime,
\begin{equation*}
\left.
\left\langle f_n^2\right\rangle_{\rm ev}
\right|_{\eta_1=\eta_2=0}
\propto
\int_0^{2\pi}d\Delta\phi\,
e^{in\Delta\phi}
\left\langle
\delta {\cE}_{\rm ev}(\mathbf{n}_1)\,
\delta {\cE}_{\rm ev}(\mathbf{n}_2)
\right\rangle_{\rm ev} .
\label{eq:EECstressTensor}
\end{equation*}
This is a mean squared, energy-weighted anisotropic flow coefficient, which has not yet been reported experimentally for the soft sector. We refer to~\cite{Barata:2023zqg,Barata:2025zku,Barata:2025uxp} for related constructions using azimuthal EECs in the hard sector.

We now
relate the final-state stress-energy correlator to its initial-condition
counterpart. The initial matching surface is at proper
time $\tau_0$, and we keep final-state observables at pseudorapidity $\eta=0$. We further assume longitudinal
boost invariance for the collision, so that the initial density profile
depends only on the transverse coordinate $\bm x$.

Focusing on the geometric, event-by-event contribution to the large-angle EEC, for each event we write
\begin{equation}
 T_{\rm ev}^{\tau\tau}(\tau_0,\bm x) = e_{\rm ev}(\bm x)
 = 
 \bar e ({\bf x}) +\delta e_{\rm ev}(\bm x),
 \label{eq:initialEnergyFluctuation}
\end{equation}
where $\bm x$ is a transverse coordinate on the early matching surface at $\tau_0$, $\bar e = \langle e_{\rm ev}(\bm x) \rangle_{\rm ev}$, and $\left\langle\delta e_{\rm ev}(\bm x)\right\rangle_{\rm ev}=0$. 
As we now see, the subsequent construction is particularly clean and elegant if one uses EECs, which enable us to formulate the matching problem entirely through the stress-energy tensor and the linear light-ray projection, and avoid invoking exclusive hadron distributions whose information content cannot be captured by $T^{\mu\nu}$ alone. 

To proceed, we take
the energy density to be the only independent field specifying
the initial stress tensor $T^{\mu\nu}$. We then write the event-wise
map from the early energy-density profile to the energy-flow
Fourier coefficients as the formal composition ($n>0$)
\begin{equation}
 F_{n,{\rm ev}}
 =
 \left(\mathcal P_n\circ\mathcal L\circ\mathcal H\right)
 [e_{\rm ev}]\, .
 \label{eq:fullResponseMap}
\end{equation}
Here, $\mathcal H$ denotes the full nonlinear map from the
initial density profile to the event-wise stress tensor field,
\begin{equation}
 T_{\rm ev}^{\mu\nu}(y)
 \equiv
 \left\langle
 \widehat T^{\mu\nu}(y)
 \right\rangle_{{\rm fixed}\,{\rm ev}}
 =
 \bigl[\mathcal H[e_{\rm ev}]\bigr]^{\mu\nu}(y)\, .
 \label{eq:dynamicalMap}
\end{equation}
The subsequent asymptotic energy-flux (light-ray) projection, $\mathcal L$, and the final angular projection, $\mathcal P_n$, are instead linear maps in their functional arguments.

Now, the key step in our construction is to characterize the nonlinear
dependence of the map $\mathcal H$ through a functional Taylor
expansion around the smooth profile $\bar e$:
\begin{align}
\nonumber &T_{\rm ev}^{\mu\nu}(y)
 =
 T_{(0)}^{\mu\nu}(y)\\
 &~~+
 \sum_{p\geq1}\frac{1}{p!}
 \int d^{2p}X\,
 \bigl[H_{(p)}(X)\bigr]^{\mu\nu}(y)
 \prod_{a=1}^{p}\delta e_{\rm ev}(\bm x_a),
 \label{eq:stressFunctionalTaylor}
\end{align}
where $X=(\bm x_1,\ldots,\bm x_p)$ denotes a set of transverse
points on the initial matching surface, and
$d^{2p}X\equiv\prod_{a=1}^{p}d^2x_a$. The background field and
response kernels are defined by
\begin{align}
 T_{(0)}^{\mu\nu}(y)
 &\equiv
 \bigl[\mathcal H[\bar e]\bigr]^{\mu\nu}(y),
 \nonumber\\
 \bigl[H_{(p)}(X)\bigr]^{\mu\nu}(y)
 &\equiv
 \left.
 \frac{\delta^p
 \bigl[\mathcal H[e]\bigr]^{\mu\nu}(y)}
 {\delta e(\bm x_1)\cdots\delta e(\bm x_p)}
 \right|_{e=\bar e}.
 \label{eq:stressResponseFunctions}
\end{align}
For hydrodynamic evolution, $H_{(1)}$ is the retarded propagator of the equations linearized about $\bar e$, giving the usual linear-response approximation.
Applying the energy-flux and angular projections, we
obtain the compact result
\begin{equation}
    F_{n,{\rm ev}}
 =
 \sum_{p\geq1}\frac{1}{p!}
 \int d^{2p}X\,
  h_n^{(p)}(X)
 \prod_{a=1}^{p}\delta e_{\rm ev}(\bm x_a),
\end{equation}
where we encapsulate the dependence on the initial transverse coordinates in the perturbation kernels 
\begin{align}
\nonumber h_n^{(p)}&(X)
 = \left(\mathcal P_n\circ\mathcal L\right)
 \biggl(\bigl[H_{(p)}(X)\bigr]^{\mu\nu}  (y) \biggr )
 =
 \int_0^{2\pi}d\phi\,e^{in\phi} \\
&\hspace{-.9cm}
\times \lim_{R\to\infty}R^2
 \int_{0}^{\infty}d t\,
 n_i(\phi)\,
 \bigl[H_{(p)}(X)\bigr]^{0i}
 \!\left(t,R\, \mathbf{n}(\phi)\right)\, .
 \label{eq:projectedResponseFunctions}
\end{align}
The contribution with $p=0$ vanishes for $n>0$ because the
background profile, and hence its evolved energy flow, is
azimuthally symmetric. Consequently, the corresponding harmonic of the geometric
energy-energy correlator, equivalently the mean-squared Fourier
coefficient, is
\begin{align}
 &\left\langle f_n^2 \right\rangle_{\rm ev}
 \equiv
 \left\langle
 F_{n,{\rm ev}}F_{n,{\rm ev}}^*
 \right\rangle_{\rm ev}
 \nn
 &{\hspace{-.1cm}}=\int d^2{\bm x}_1\,d^2{\bm x}_2\,
 h_n^{(1)}(\bm x_1)
 h_n^{(1)*}(\bm x_2)
 \left\langle
 \delta e_{\rm ev}(\bm x_1)
 \delta e_{\rm ev}(\bm x_2)
 \right\rangle_{\rm ev}
\nn
&
 +\mathcal O\!\left[(\delta e_{\rm ev})^3\right].
 \label{eq:linearLinearEEC}
\end{align}
We emphasize that no assumption of linear dynamics has been made so far, and that we cannot tell a priori whether Eq.~\eqref{eq:linearLinearEEC} should be truncated at first order ($p=1$).

However, taking insight from the usual geometry-to-flow picture of heavy-ion collisions, we note that Eq.~\eqref{eq:linearLinearEEC} has a striking resemblance to the standard mean squared eccentricity of the initial density field \cite{Blaizot:2014nia}. Consider the multipole moment $\epsilon_{\ell,m}= \int d^2{\bm r} \,r^\ell e^{im\varphi} \delta e({\bm r})$, such that 
\begin{equation}
     \hspace{-.2 cm}\langle \epsilon_{n,n} \epsilon^*_{n,n} \rangle_{\rm ev} = \int_{\bm r_1 , \bm r_2} r_1^n \,r_2^n \,e^{in(
    \varphi_1-\varphi_2 
    )  }  \langle \delta e({\bm r}_1)\delta e({\bm r}_2)  \rangle_{\rm ev} \, . 
    \label{eq:Eecce}
\end{equation}
Now, much as the standard dimensionless $V_n$ is correlated with the dimensionless initial-energy eccentricity, the energy-weighted $f_n$ will also be correlated with the energy-weighted anisotropy of Eq.~(\ref{eq:Eecce}), since both quantities arise from the same dynamical expansion of the system and at particle level differ solely by a transverse-energy weight for the correlated hadrons. Therefore, if $h^{(1)}_n$ in Eq.~\eqref{eq:linearLinearEEC} admits the form $ h^{(1)}_n({\bm r}) \propto r^n e^{in\varphi}$,
then truncating Eq.~\eqref{eq:linearLinearEEC} becomes equivalent to a linear response to eccentricity.

As shown in the Supplement, this has solid motivation. Rotational
covariance of the Fourier coefficients implies
\begin{equation}
 h_n^{(1)}(r,\varphi)
 =
 e^{in\varphi}\rho_n(r),
 \label{eq:angularResponse}
\end{equation}
whereas the radial dependence can be expanded in modes with leading
term proportional to $r^n$ (e.g. an orthonormal basis of polynomials). The Teaney-Yan cumulant expansion \cite{Teaney:2010vd} motivates retaining this lowest
mode, such that substituting $ h_n^{(1)}(r,\varphi)
 \simeq
 g_{n0}\,r^n e^{in\varphi}
 \label{eq:leadingRadialResponse}
$,  where $g_{n0}$ is a response coefficient, into
Eq.~\eqref{eq:linearLinearEEC} gives (for $n\geq 2$)
\begin{equation}
 \left\langle f_n^2\right\rangle_{\rm ev}
 \simeq
 |g_{n0}|^2
 \bigl\langle
 \left|\epsilon_{n,n}\right|^2
 \bigr\rangle_{\rm ev},
 \label{eq:energyWeightedEccentricityResponse}
\end{equation}
up to higher-radial modes and nonlinear corrections.

Some remarks are in order. At linear order, Eq.~\eqref{eq:energyWeightedEccentricityResponse}
has the same structure as the mode-by-mode construction of
Floerchinger and Wiedemann~\cite{Floerchinger:2013rya}, which decomposes
the initial fluctuating field into Fourier--Bessel modes, each
propagating independently to the final hadronic spectrum. Our
construction instead organizes the radial dependence in a basis generated by
$r^n,r^{n+2},r^{n+4},\ldots$. This is merely a change of basis, but it allows us to physically motivate the
truncation of Eq.~(\ref{eq:linearLinearEEC}) to the lowest order. An analogous linear-response stress tensor construction
underlies the pre-equilibrium framework
K{\o}MP{\o}ST~\cite{Kurkela:2018vqr}, albeit without truncation in the radial modes. A study of the nuclear collision through linear response of a unified stress-energy-tensor evolution was also recently explored in \cite{Kirchner:2024woh}.

Once more, while
Ref.~\cite{Floerchinger:2013rya} terminates in a number-weighted hadronic
correlator obtained after freeze-out, here we match to the
asymptotic energy flux, such that both endpoints
are expressed in terms of the same QCD operator:
\begin{align*}
 &\int d\Delta\phi\,
 e^{in\Delta\phi}
 \left\langle
 \delta\mathcal E_{\rm ev}(\phi_1)
 \delta\mathcal E_{\rm ev}(\phi_2)
 \right\rangle_{\rm ev}
 \nonumber\\
\nonumber &\propto
 \int  d^2{\bm r}_1\,d^2{\bm r}_2\,
 r_1^n r_2^n e^{in(\varphi_1-\varphi_2)}
 \left\langle
 \delta T_{\rm ev}^{\tau\tau}(\tau_0,\bm r_1)
 \delta T_{\rm ev}^{\tau\tau}(\tau_0,\bm r_2)
 \right\rangle_{\rm ev} \\
 & ~~~~~~~~~~~~~+ \mathcal{O}\left [(\delta T_{\rm ev}^{\tau\tau})^3 , r^{2n+2} (\delta T_{\rm ev}^{\tau\tau})^2 \right]\,.
 \label{eq:stressTensorMatching}
\end{align*}
This fulfills our initial promises. First, the endpoints of this matching are formulated within the common language of the conserved stress-energy tensor. Second, the eccentricity scaling \textit{emerges} in our construction as the leading order in an expansion where higher radial moments and nonlinear response terms appear explicitly as systematically identifiable corrections, following the Teaney-Yan double-expansion logic. Making this hierarchy manifest is a stepping stone toward constructing a power-counting and operator-level organization of the dynamics of these correlators in a forthcoming EFT formulation.

We now turn to the second half of the matching and connect the
early-time stress tensor to the many-body structure of the incoming
nuclear ground states. This step follows in spirit
Refs.~\cite{Giacalone:2023hwk,Duguet:2025hwi,Mehrabpour:2025ogw,Mehrabpour:2026yuc}, where
correlation functions characterizing the initial collision field are
related to density correlations in the incoming nuclei.

For an ultrarelativistic nucleus, the stress tensor is dominated by
its light-front component, $T^{++}~(T^{--})$ for the right-moving (left-moving) nucleus. In such a limit, we encode all the information in a two-dimensional thickness,
\begin{equation}
 t^{+}_{{\rm ev}}(\bm x)
 \ \propto\
 \int d x^-\,T_{{\rm ev}}^{
++
 }(x^-,\bm x),
 \label{eq:lightFrontThickness}
\end{equation}
and analogously for $t^-$, where the proportionality depends on normalization conventions and the physical problem under consideration.
In general, at a resolution for which nucleons provide the appropriate
low-energy degrees of freedom, the thickness function
can be represented as a one-body density operator,
\begin{equation}
 \widehat t(\bm x)
 =
 \sum_{i=1}^{A}
 \delta(\bm x-\widehat{\bm x}_i),
 \label{eq:thicknessOperator}
\end{equation}
where $\widehat{\bm x}_i$ is the transverse position of nucleon
$i$. This is, then, the core object underlying all implementations of high-energy scattering processes with nuclei.

A collision of light-front sheets constructed as an effective theory of QCD can be achieved within the Color Glass Condensate (CGC) formalism~\cite{McLerran:1993ni,McLerran:1993ka,McLerran:1994vd}, where the thickness function plays the role of the local variance of color charge fluctuations. The CGC postulates a separation of scales entailing that
large-$x$ degrees of freedom in the boosted nucleus act as approximately static color
sources of small-$x$ gluon fields, with the local density of the latter characterized by a saturation momentum $Q_s$. After coarse-graining over partonic
transverse scales, the slowly varying dependence on the nuclear
geometry can be parametrized as
\begin{equation*}
 (Q^\pm_{s})^{2}(\bm x)
 \simeq \mathcal C_0^\prime \, t^\pm_{{\rm ev}}(\bm x) + \ldots,
 \label{eq:saturationThickness}
\end{equation*}
for some matching coefficient $\mathcal C_0^\prime$, and where the ellipsis denotes subleading short-range structures,
transverse-gradient corrections, or higher-body operators.
We regard this as a leading coarse-grained matching relation. Constructing the corresponding systematic operator expansion in an EFT is left for future work. Note that such corrections are implemented in phenomenological applications through models such as IP-Sat~\cite{Schenke:2020mbo}, including possible nonlinear QCD evolution effects \cite{Mantysaari:2024qmt,Mantysaari:2025tcg}.

Immediately after the collision, the initial energy density
can be factorized into contributions associated with the gluon
distributions of the two incoming nuclei
\cite{Krasnitz:1999wc,Lappi:2006hq,Albacete:2018bbv}, which motivates
the leading local map
\begin{equation*}
 e_{\rm ev}(\bm x)
 \propto
 (Q^+_{s})^{2}(\bm x)(Q^-_{s})^{2}(\bm x)
 \simeq
 \mathcal C_0^2\,
 t^+_{{\rm ev}}(\bm x)t^-_{{\rm ev}}(\bm x),
 \label{eq:GlasmaThicknessMatching}
\end{equation*}
for some other $\mathcal C_0$. Expanding around the common mean nuclear profile,
$
 t_{{\rm ev}}(\bm x)
 =
 \bar t ({\bm x})+\delta t_{{\rm ev}}(\bm x)$, with $\langle\delta t_{\rm ev}(\bm x)\rangle_{\rm ev}=0$, and considering symmetric collisions with the two nuclei having statistically independent, albeit identical, fluctuation spectra, the previous relation becomes \cite{Zhou:2025bwu}
\begin{align}
\nonumber   \bigl\langle \delta e_{\rm ev}(\bm x_1)
 & \delta e_{\rm ev}(\bm x_2)
 \bigr\rangle_{\rm ev}  \simeq \\
 & 2 \mathcal{C}_0^4 \, \bar t (\bm x_1)\bar t(\bm x_2)\,
  \langle \delta t_{\rm ev}(\bm x_1) \delta t_{\rm ev} (\bm x_2) \rangle_{\rm ev}\, ,
 \label{eq:linearEnergyThicknessCovariance}
\end{align}
up to $\mathcal O(\delta t^4)$. We thus express energy density correlations in terms of nuclear density correlations. 

As also detailed in the Supplement, within the leading-response approximation used above,
$h_n^{(1)}(r,\varphi)\simeq g_{n0}r^ne^{in\varphi}$, reducing to the nucleonic operators yields
\begin{equation}
     \left\langle f_n^2\right\rangle_{\rm ev}
 \propto
 \left\langle
 \widehat{\mathcal Q}^{[\bar t]}_n
 \left(\widehat{\mathcal Q}^{[\bar t]}_n\right)^\dagger
 \right\rangle_\Psi
 +\ldots ,
\end{equation}
expressed thus in terms of the \emph{thickness-weighted} multipole
\begin{align}
 \widehat{\mathcal Q}^{[\bar t]}_n
 \equiv
 \int d^2x\,\bar t(r)r^ne^{in\varphi}\hat t(\bm x)
 =\sum_{i=1}^{A}\bar t(r_i)r_i^ne^{in\varphi_i},
\end{align}
and where $\Psi$ denotes the nuclear ground state. The thickness weight does not generate new angular structures, but only
reshuffles coefficients within the same radial tower. Indeed, writing
$\bar t(r)=\sum_{j\geq0}b_jr^{2j}$ and
$\rho_n(r)=\sum_{\ell\geq0}g_{n\ell}r^{n+2\ell}$ gives
\begin{equation}
 \bar t(r)\rho_n(r)
 =\sum_{j,\ell\geq0} b_j g_{n\ell}\,r^{n+2(j+\ell)}.
\end{equation}
Applying the same dynamically motivated radial truncation to the complete response kernel $\bar t(r)\rho_n(r)$, we retain only its lowest radial contribution with $j=\ell=0$.
At this order, the contribution of $\widehat{\mathcal Q}^{[\bar t]}_n$
reduces to that of the bare multipole operator $\widehat{\mathcal Q}_n$
derived in \cite{Duguet:2025hwi}, with the prefactor  absorbed into the full response coefficient. This leads to
\begin{equation*}
 \left\langle f_n^2\right\rangle_{\rm ev}
 \propto
 \bigl \langle 
 \widehat{\mathcal Q}_n
 \widehat{\mathcal Q}_n^\dagger
 \bigr \rangle_\Psi
 +\ldots \,.
 \label{eq:EECNuclearOperator}
\end{equation*}
This completes the desired initial-to-final-state mapping.

Therefore, to lowest order in the dynamical response (or density perturbation, $\delta e$), 
to leading order in the thickness fluctuation, $\delta t_{\rm ev}$, and to leading order in the  
 matching of the light-front QCD stress tensor to a thickness function (or saturation scale), the mean squared Fourier coefficient associated with the azimuthal projection of the EEC, $\langle f_{n}^2\rangle_{\rm ev}$, provides a measure of the expectation of a multipole-operator correlator evaluated in the nuclear ground state, $\langle 
 \widehat{\mathcal Q}_n
 \widehat{\mathcal Q}_n^\dagger
 \rangle_\Psi$. This is the starting point for an EFT formulation
linking stress-tensor correlators in the incoming nuclei to those in the final state, encoding short-distance QCD dynamics in matching coefficients and consistently organizing corrections to the leading nuclear operators.


For the future, it would be of great interest to generalize this work to higher-point energy correlators, expanding on the existing analyses of hard probes~\cite{Bossi:2024qho,Barata:2025fzd,Budhraja:2025ulx}. In the language of flow analyses, projections leading to the correlator $\langle F_{n,\rm ev} F_{n,\rm ev}^* \delta F_{0,\rm ev} \rangle_{\rm ev}$ give the energy-flow analogue of the correlation between $|V_n|^2$ and the mean transverse momentum \cite{Bozek:2016yoj,Giacalone:2019pca,Schenke:2020uqq,Giacalone:2020dln,Giacalone:2020awm,Jia:2021wbq,Bally:2021qys,Zhao:2024lpc,Hagino:2025vxe,Zhang:2025zrm}, and would probe three-point correlations and non-Gaussian properties of the initial states. Such a discussion finds a parallel in the literature on asymptotic detectors~\cite{Chen:2022swd}. Moreover, the extension to detectors of other conserved currents~\cite{Hofman:2008ar}, such as electric charge or baryon number, might further help in probing the associated initial-state fluctuations, see e.g.~\cite{Carzon:2019qja,Sousa:2025oqf}.

More broadly, and intriguingly, the structure uncovered in this work suggests a simple interpretation of the collision process as a scale-dependent filter. Short-distance information is scrambled and absorbed into response functions and matching coefficients, whereas sufficiently long-wavelength correlations can be transmitted through the evolution and remain encoded in asymptotic observables. Collisions involving atomic nuclei, which present a variety of emergent long-range collective phenomena, provide an especially striking realization of this principle. However, exploring this viewpoint beyond the specific setting of nuclear structure may reveal a broader role for energy-flow observables as probes of long-wavelength physics in different theoretical setups and many-body quantum states.

\bigskip
G.G. especially thanks Thomas Duguet for discussions that helped motivate this project. We acknowledge discussions with Matt Luzum, Jean-Yves Ollitrault, Petja Paakkinen, Yiyang Peng, Tom Reichert, Andrey Sadofyev, Wilke van der Schee, Huichao Song, Adam Takacs, and Jiangming Yao.

\bibliographystyle{bibstyle.bst}

\bibliography{references.bib}
\clearpage

\appendix
\section*{supplemental material}


\section{1. Energy-flow harmonics away from midrapidity}
\label{app:finite-rapidity}

In this appendix we generalize Eq.~\eqref{eq:EECharmonics} to arbitrary
pseudorapidities and to finite pseudorapidity windows.  The powers of
$\cosh\eta$ depend on the measure and on whether the observable is weighted
by energy or by transverse energy.  We therefore keep these choices explicit
throughout.  Unless stated otherwise, $\eta$ denotes the pseudorapidity, $\eta=-\ln\tan\frac{\theta}{2}$.
In particular, the solid angle integration measure is
\begin{equation}
  d\Omega=\left|\sin\theta\,d\theta\right|d\phi
  =\frac{d\eta\,d\phi}{\cosh^2\eta}\, .
  \label{eq:app-solid-angle-jacobian}
\end{equation}

The energy-flow operator event-wise expectation value can be written as $ \mathcal E_{\rm ev}(\eta,\phi)
  \equiv \frac{dE}{d\Omega}\big|_{\rm ev}$. On the other hand, we define the transverse-energy density per unit
pseudorapidity and azimuth as 
\begin{equation}
  \mathcal F_{\rm ev}(\eta,\phi)
  \equiv
  \frac{dE_T}{d\eta\,d\phi}\bigg|_{\rm ev},
  \qquad
  E_T=E\sin\theta=\frac{E}{\cosh\eta}.
  \label{eq:app-F-definition}
\end{equation}
Using Eq.~\eqref{eq:app-solid-angle-jacobian} and
$dE/d\eta d\phi=\cosh\eta\,dE_T/d\eta d\phi$ gives the exact kinematic relation $\mathcal E_{\rm ev}(\eta,\phi)
  =\cosh^3\eta\,\mathcal F_{\rm ev}(\eta,\phi)$. Thus the factor $\cosh^3\eta$ contains one power from converting $E_T$ to
$E$ and two powers from converting a density per $d\eta\,d\phi$ to a density
per $d\Omega$.  Notice that we have not required boost invariance. When applied directly to a non-interacting asymptotic multiparticle state, the above distributions reduce to
\begin{align}
  \mathcal F_{\rm ev}(\eta,\phi)
  &=\sum_{i\in{\rm ev}}
    \frac{E_i}{\cosh\eta_i}
    \delta(\eta-\eta_i)\delta(\phi-\phi_i),
\nn
  \mathcal E_{\rm ev}(\eta,\phi)
  &=\sum_{i\in{\rm ev}}
    E_i\cosh^2\eta_i\,
    \delta(\eta-\eta_i)\delta(\phi-\phi_i).
  \label{eq:app-particle-E}
\end{align}
Here we use notation
$\mathcal E_{\rm ev}$ and $\mathcal F_{\rm ev}$ for the energy
distributions of an individual final-state particle configuration.
Their conditional averages at fixed event geometry define the
event-wise fields used in the main text and in the following.
As in the main text, we neglect the fixed-geometry contribution
to the EEC, retaining only its geometric component.

Two related two-point densities can thus be introduced:
the correlator $\mathcal G_2^{\Omega}$ of energy flow per unit solid angle and the
correlator $\mathcal G_2^{T}$ of transverse energy per $d\eta\,d\phi$.  Their kinematic
relations are
\begin{align}
  \mathcal G_2^{\Omega}(\mathbf{n}_1,\mathbf{n}_2)
  &=\cosh^3\eta_1\cosh^3\eta_2\,
    \mathcal G_2^{T}(\mathbf{n}_1,\mathbf{n}_2).
  \label{eq:app-three-G-E-T}
\end{align}

At each pseudorapidity, we define the dimensionful transverse-energy harmonics
\begin{align}
  F_{n,{\rm ev}}(\eta)
  &\equiv\int_0^{2\pi}d\phi\,
  e^{in\phi}\mathcal F_{\rm ev}(\eta,\phi),
  \notag\\[-0.2em]
  F_{-n,{\rm ev}}(\eta)&=F_{n,{\rm ev}}^*(\eta).
  \label{eq:app-Fn-eta-definition}
\end{align}
Consequently,
\begin{align}
  \mathcal F_{\rm ev}(\eta,\phi)
  &=\frac{1}{2\pi}\sum_{n\in\mathbb Z}
    F_{n,{\rm ev}}(\eta)e^{-in\phi},
  \nn
  \mathcal E_{\rm ev}(\eta,\phi)
  &=\frac{\cosh^3\eta}{2\pi}\sum_{n\in\mathbb Z}
    F_{n,{\rm ev}}(\eta)e^{-in\phi}.
  \label{eq:app-E-Fourier-eta}
\end{align}
Writing $\delta F_{n,{\rm ev}}(\eta)
  \equiv F_{n,{\rm ev}}(\eta)
   -\big\langle F_{n,{\rm ev}}(\eta)\big\rangle_{\rm ev}$,
rotational invariance of the event ensemble implies
$\langle F_{n,{\rm ev}}(\eta)\rangle_{\rm ev}=0$ for $n\ne0$ and
\begin{equation}
  \left\langle
    \delta F_{n,{\rm ev}}(\eta_1)
    \delta F_{m,{\rm ev}}(\eta_2)
  \right\rangle_{\rm ev}
  =\delta_{m,-n}\,C_n^F(\eta_1,\eta_2),
  \label{eq:app-rotational-selection-rule}
\end{equation}
where
\begin{equation}
  C_n^F(\eta_1,\eta_2)
  \equiv\left\langle
    \delta F_{n,{\rm ev}}(\eta_1)
    \delta F_{n,{\rm ev}}^*(\eta_2)
  \right\rangle_{\rm ev}.
  \label{eq:app-CnF-definition}
\end{equation}
For $n=0$, this is the covariance of the transverse energies per unit
pseudorapidity.  For $n>0$, the subtractions in
Eq.~\eqref{eq:app-CnF-definition} vanish in a rotationally invariant
ensemble.  The covariance obeys
\begin{equation}
  C_n^F(\eta_1,\eta_2)^*=C_n^F(\eta_2,\eta_1).
  \label{eq:app-Cn-hermiticity}
\end{equation}
It need not be real when $\eta_1\ne\eta_2$, as an imaginary part can encode a rapidity-dependent rotation of the event plane.  

Using Eqs.~\eqref{eq:app-E-Fourier-eta} and
\eqref{eq:app-rotational-selection-rule}, the geometric contribution to the
EEC at two arbitrary pseudorapidities is
\begin{equation}
  \begin{aligned}
  \mathcal G_{2}^{\Omega,{\rm geom}}(\mathbf{n}_1,\mathbf{n}_2)
  &=\frac{\cosh^3\eta_1\cosh^3\eta_2}{(2\pi)^2}
  \sum_{n\in\mathbb Z}
   C_n^F(\eta_1,\eta_2)e^{-in(\phi_1-\phi_2)}\,.
  \end{aligned}
  \label{eq:app-G2-finite-eta}
\end{equation}
This is the finite-pseudorapidity counterpart of
Eq.~\eqref{eq:EECharmonics}.  The sign in the exponential follows from the
Fourier convention in Eq.~\eqref{eq:app-Fn-eta-definition}.   
If the relevant
reflection symmetry makes $C_n^F$ real, Eq.~\eqref{eq:app-G2-finite-eta}
can be written as
\begin{align}
  \mathcal G_{2}^{\Omega,{\rm geom}}(\mathbf{n}_1,\mathbf{n}_2)
  &=\frac{\cosh^3\eta_1\cosh^3\eta_2}{(2\pi)^2}
  \bigg[
    C_0^F(\eta_1,\eta_2)
  \nn &+2\sum_{n=1}^{\infty}C_n^F(\eta_1,\eta_2)
      \cos(n\Delta\phi)
  \bigg],
  \label{eq:app-G2-finite-eta-cosine}
\end{align}
with $\Delta\phi\equiv\phi_1-\phi_2$. A direct projection gives
\begin{align}
  &\int_0^{2\pi}d\phi_1\int_0^{2\pi}d\phi_2\,
   e^{in(\phi_1-\phi_2)}
   \mathcal G_{2}^{\Omega,{\rm geom}}(\mathbf{n}_1,\mathbf{n}_2)
  \notag\\
  &\hspace{2cm}=
  \cosh^3\eta_1\cosh^3\eta_2\,
  C_n^F(\eta_1,\eta_2).
  \label{eq:app-double-projection}
\end{align}
Equivalently, when rotational invariance makes the correlator a function of
$\Delta\phi$ alone,
\begin{align}
  C_n^F(\eta_1,\eta_2)
  &=\frac{2\pi}{\cosh^3\eta_1\cosh^3\eta_2}
   \int_0^{2\pi}d\Delta\phi\,
  \nn
  &\hspace{0cm}\times e^{in\Delta\phi}
   \mathcal G_{2}^{\Omega,{\rm geom}}
   (\eta_1,\eta_2;\Delta\phi)\,,
  \label{eq:app-single-projection}
\end{align}
so that at equal pseudorapidities, using
$F_{n,{\rm ev}}(\eta)=f_{n,{\rm ev}}(\eta)
e^{in\Psi_{n,{\rm ev}}(\eta)}$ for $n>0$, we find
\begin{equation}
  C_n^F(\eta,\eta)
  =\left\langle |F_{n,{\rm ev}}(\eta)|^2\right\rangle_{\rm ev}
  =\left\langle f_{n,{\rm ev}}^2(\eta)\right\rangle_{\rm ev}\, ,
  \label{eq:app-fn2-equal-eta}
\end{equation}
and Eq.~\eqref{eq:app-single-projection} becomes the desired finite-rapidity formula
\begin{equation}
  \begin{aligned}
  \left\langle f_n^2(\eta)\right\rangle_{\rm ev}
  &=\frac{2\pi}{\cosh^6\eta}
   \int_0^{2\pi}d\Delta\phi\,e^{in\Delta\phi}
  \nn
  &\times\mathcal G_{2}^{\Omega,{\rm geom}}
   (\eta,\eta;\Delta\phi)\, .
  \end{aligned}
  \label{eq:app-fn2-finite-eta}
\end{equation}
For two different pseudorapidities one measures instead
\begin{equation}
  C_n^F(\eta_1,\eta_2)
  =\left\langle
    f_n(\eta_1)f_n(\eta_2)
    e^{in[\Psi_n(\eta_1)-\Psi_n(\eta_2)]}
  \right\rangle_{\rm ev},
  \label{eq:app-longitudinal-decorrelation}
\end{equation}
which retains both magnitude decorrelation and event-plane twist along the
longitudinal direction.

A generalization of the transverse-energy harmonic in the main text is obtained by choosing a real acceptance weight $w_A(\eta)$ supported in a window $A$ and defining
\begin{align}
  \mathcal F_{A,{\rm ev}}(\phi)
  &\equiv\int_A d\eta\,w_A(\eta)
     \mathcal F_{\rm ev}(\eta,\phi),
  \nn
  F_{n,A,{\rm ev}}
  &\equiv\int_A d\eta\,w_A(\eta)F_{n,{\rm ev}}(\eta).
  \label{eq:app-window-Fn}
\end{align}
Two windows $A$ and $B$ can be used to impose a pseudorapidity gap, and their harmonic covariance is
\begin{align}
  C_{n;AB}^F
  &\equiv\left\langle
    \delta F_{n,A,{\rm ev}}
    \delta F_{n,B,{\rm ev}}^*
  \right\rangle_{\rm ev}
  \notag\\
  &=\int_A d\eta_1\int_B d\eta_2\,
    w_A(\eta_1)w_B(\eta_2)
    C_n^F(\eta_1,\eta_2).
  \label{eq:app-window-Cn}
\end{align}
For $A=B$ and $n>0$ this is
$C_{n;AA}^F=\langle|F_{n,A,{\rm ev}}|^2\rangle_{\rm ev}$. Combining Eq.~\eqref{eq:app-window-Cn} with
Eq.~\eqref{eq:app-double-projection} gives 
\begin{align}
  C_{n;AB}^F
  &=\int_A d\eta_1\int_B d\eta_2
    \int_0^{2\pi}d\phi_1\int_0^{2\pi}d\phi_2\,
    \frac{w_A(\eta_1)w_B(\eta_2)}
    {\cosh^3\eta_1\cosh^3\eta_2}
  \nn&\times e^{in(\phi_1-\phi_2)}
    \mathcal G_{2}^{\Omega,{\rm geom}}(\mathbf{n}_1,\mathbf{n}_2).
  \label{eq:app-window-extraction}
\end{align}
Each factor $1/\cosh^3\eta_i$ converts one insertion of energy flow per
solid angle into transverse energy per $d\eta_i\,d\phi_i$.  Equivalently,
one may first define the window-integrated transverse-energy correlator
\begin{align}
  \mathcal G_{2;AB}^{T,{\rm geom}}(\phi_1,\phi_2)
  &\equiv\int_A d\eta_1\int_B d\eta_2\,
  \frac{w_A(\eta_1)w_B(\eta_2)}
       {\cosh^3\eta_1\cosh^3\eta_2}
  \nn&\times
  \mathcal G_{2}^{\Omega,{\rm geom}}(\mathbf{n}_1,\mathbf{n}_2),
  \label{eq:app-window-GT}
\end{align}
for which
\begin{equation}
  \mathcal G_{2;AB}^{T,{\rm geom}}(\phi_1,\phi_2)
  =\frac{1}{(2\pi)^2}\sum_{n\in\mathbb Z}
   C_{n;AB}^F e^{-in(\phi_1-\phi_2)}\,. 
  \label{eq:app-window-GT-Fourier}
\end{equation}

For an exactly boost-invariant event, the
transverse-energy pattern is independent of pseudorapidity over the region of interest, i.e. $\mathcal F_{\rm ev}(\eta,\phi)=\mathcal F_{\rm ev}(\phi)$ and $
  F_{n,{\rm ev}}(\eta)=F_{n,{\rm ev}}$.
In this case we then have
\begin{equation}
  C_n^F(\eta_1,\eta_2)=
  \begin{cases}
  \big\langle(\delta F_0)^2\big\rangle_{\rm ev},&n=0,\\[0.2em]
  \big\langle f_{|n|}^2\big\rangle_{\rm ev},&n\ne0,
  \end{cases}
  \label{eq:app-Cn-boost-invariant}
\end{equation}
and Eq.~\eqref{eq:app-G2-finite-eta} becomes
\begin{align}
  \mathcal G_{2}^{\Omega,{\rm geom}}(\mathbf{n}_1,\mathbf{n}_2)
  &=\frac{\cosh^3\eta_1\cosh^3\eta_2}{(2\pi)^2}
  \nn 
  &\hspace{-1 cm}\times \bigg[\big\langle(\delta F_0)^2\big\rangle_{\rm ev}
  +\sum_{n\in\mathbb Z\setminus\{0\}}
  \big\langle f_{|n|}^2\big\rangle_{\rm ev}
  e^{-in\Delta\phi}\bigg].
  \label{eq:app-G2-boost-invariant}
\end{align}
Setting $\eta_1=\eta_2=0$ recovers Eq.~\eqref{eq:EECharmonics}.

\section{2. Response functional and long-wavelength expansion}
\label{app:response-map}

In this Appendix we give a more explicit formulation of the response map
$\mathcal H$ introduced in the main text, and we clarify the two
approximations on which the main analysis rests: an expansion in the
amplitude of the event-by-event fluctuations around a smooth background,
and, within the linear term of that expansion, a long-wavelength truncation
to the lowest radial mode.

Let $\Sigma_0$ denote the initial matching surface at proper time $\tau_0$.
At vanishing conserved-charge densities and imposing boost invariance, the
macroscopic state on this surface is described by a set of independent
fields
\begin{equation}
 \Phi_A(\tau_0,\bm x)
 =
 \bigl\{
 e,\,u^x,\,u^y,\,\Pi,\,\pi^{xx},\,\pi^{xy},\ldots
 \bigr\}(\tau_0,\bm x),
 \label{app:field-vector}
\end{equation}
where the index $A$ labels the independent components after imposing the
normalization of the velocity, the transversality and tracelessness
conditions on the shear tensor, and a choice of hydrodynamic frame;
equivalently, one may take a set of independent components of $T^{\mu\nu}$
itself as the fundamental variables.  In a pre-equilibrium description the
entries of Eq.~\eqref{app:field-vector} are simply replaced by a complete
set of macroscopic variables appropriate to that stage.

For fixed initial data, the expectation value of the stress tensor at a
later space-time point $y$ is a nonlinear functional,
\begin{equation}
 T^{\mu\nu}_{\rm ev}(y)
 =
 \bigl[\mathcal H[\Phi_{\rm ev}(\tau_0)]\bigr]^{\mu\nu}(y).
 \label{app:general-H}
\end{equation}
Here $T^{\mu\nu}_{\rm ev}$ denotes the conditional, event-wise mean stress
tensor; stochastic fluctuations at fixed initial condition are not part of
$\mathcal H$, and enter the energy correlator through the fixed-condition
term of the covariance decomposition of the main text.  The functional
$\mathcal H=\mathcal H_{\rm late}\circ\mathcal H_{\rm hydro}\circ
\mathcal H_{\rm pre}$ may contain pre-equilibrium evolution, hydrodynamic
evolution, particlization, and the late hadronic stage: while the exact
dynamics of every stage may be nonlinear, the response to sufficiently
small perturbations is propagated by linearizing each stage around its
corresponding background solution.

We decompose the initial fields into an azimuthally symmetric ensemble
background and event-wise fluctuations,
\begin{equation}
 \begin{aligned}
 \Phi_{A,{\rm ev}}(\tau_0,\bm x)
 &=
 \bar\Phi_A(\tau_0,r)
 +\delta\Phi_{A,{\rm ev}}(\tau_0,\bm x),
 \\
 \left\langle\delta\Phi_{A,{\rm ev}}(\tau_0,\bm x)
 \right\rangle_{\rm ev}&=0,
 \end{aligned}
 \label{app:background-split}
\end{equation}
and the background is evolved with the full nonlinear map,
$ T_{(0)}^{\mu\nu}(y)=[\mathcal H[\bar\Phi]]^{\mu\nu}(y)$.
Assuming differentiability of $\mathcal H$ in a neighborhood of $\bar\Phi$,
its dependence on the initial perturbations admits the functional expansion
\begin{align}
 T^{\mu\nu}_{\rm ev}(y)
 ={}&  T_{(0)}^{\mu\nu}(y)
 +
 \sum_{p\geq1}\frac{1}{p!}
 \int_{\Sigma_0}\prod_{a=1}^{p}d^2x_a\,
 \nonumber\\
 &\hspace{-1.5cm }\times
 \bigl [H_{(p)A_1\cdots A_p}
 (\bm x_1,\ldots,\bm x_p)\bigr ]^{\mu\nu} (y)
 \prod_{a=1}^{p}
 \delta\Phi_{A_a,{\rm ev}}(\tau_0,\bm x_a),
 \label{app:functional-expansion}
\end{align}
where repeated field indices are summed and
\begin{align}
 &\bigl [H_{(p)A_1\cdots A_p}
 (\bm x_1,\ldots,\bm x_p)\bigr ]^{\mu\nu} (y)
 \nonumber\\
 &\quad\equiv
 \frac{\delta^{p}\,[\mathcal H[\Phi]]^{\mu\nu}(y)}
 {\delta\Phi_{A_1}(\tau_0,\bm x_1)
  \cdots
  \delta\Phi_{A_p}(\tau_0,\bm x_p)}
 \bigg|_{\Phi=\bar\Phi}.
 \label{app:response-kernels}
\end{align}
The kernel $H_{(1)}$ is the linear propagator of perturbations on the
background, whereas the kernels with $p\geq2$ describe nonlinear mode
coupling. At linear order
Eq.~\eqref{app:functional-expansion} becomes
\begin{equation}
 \delta T^{\mu\nu}_{\rm ev}(y)
 =
 \int d^2x\,
 \bigl [H_{(1)A}(\bm x)\bigr]^{\mu\nu} (y)\,
 \delta\Phi_{A,{\rm ev}}(\tau_0,\bm x)
 +\mathcal O(\delta\Phi^2).
 \label{app:linear-general}
\end{equation}

The main text specializes to the scalar channel in which the independent
initial perturbation is the energy density, with the remaining components
of the initial stress tensor fixed, rotationally covariant functionals of
$e$. Formally, at linear order this amounts to
\begin{align}
 \delta\Phi_A(\tau_0,\bm x)
 &=
 \int d^2x'\,
 C_A(\bm x,\bm x')\,\delta e(\bm x'),
 \label{app:scalar-initialization}
 \\
\bigl [ H_{(1)e}(\bm x')\bigr ]^{\mu\nu} (y)
 &\equiv
 \int d^2x\,
 \bigl [ H_{(1)A} ({\bm x}) \bigr ]^{\mu\nu}(y)\,C_A(\bm x,\bm x'),
 \label{app:effective-scalar-green}
\end{align}
where in our initialization
$C_A(\bm x,\bm x')=\delta_{A,e}\,\delta^{(2)}(\bm x-\bm x')$, so
\begin{equation}
 \delta T^{\mu\nu}_{\rm ev}(y)
 =
 \int d^2x'\,
 \bigl [ H_{(1)e}(\bm x')\bigr ]^{\mu\nu}\,
 \delta e_{\rm ev}(\bm x')
 +\mathcal O(\delta e^2).
 \label{app:scalar-linear-response}
\end{equation}

The light-ray projection $\mathcal L$ and the angular projection
$\mathcal P_n$ are linear in the stress tensor.  Applying them to
Eq.~\eqref{app:functional-expansion} gives, for $n\neq0$, where the
symmetric background does not contribute,
\begin{align}
 F_{n,{\rm ev}}
 &=
 \sum_{p\geq1}\frac{1}{p!}
 \int\prod_{a=1}^{p}d^2x_a\,
 h^{(p)}_{n;A_1\cdots A_p}
 (\bm x_1,\ldots,\bm x_p)
 \nonumber\\
 &\times
 \prod_{a=1}^{p}
 \delta\Phi_{A_a,{\rm ev}}(\tau_0,\bm x_a),
 \label{app:projected-functional-series}
\end{align}
with
\begin{equation}
h^{(p)}_{n;A_1\cdots A_p}(X)
\equiv
(\mathcal P_n\circ\mathcal L)
\left(
 [H_{(p)A_1\cdots A_p}(X)]^{\mu\nu}(y)
\right).
 \label{app:projected-kernels}
\end{equation}
Thus, the kernel $h_n^{(1)}$ introduced in the main text is the
asymptotically projected retarded Green's function of the complete collision
evolution, containing the propagation of the initial
perturbation to the
asymptotic energy flux.  

For an azimuthally symmetric background, and as derived in the next section, in the
scalar channel one has
\begin{equation}
 h_n^{(1)}(r,\varphi)
 =
 e^{in\varphi}\rho_n(r),
 \label{app:linear-angular-kernel}
\end{equation}
as anticipated in the main text.  The angular
dependence is fixed by symmetry and all dynamical information is contained in
the radial response $\rho_n(r)$.  A basis of polynomials can always be chosen such that
\begin{align}
 \rho_n(r)
 &=\sum_{\ell\geq0}g_{n\ell}\,r^{n+2\ell}.
 \label{app:radial-kernel-expansion}
\end{align}
Therefore, expanding the response kernels in such a basis enables us to describe the final-state two-point correlator as a hierarchy of moments of multi-point correlation functions of the field $\delta e$.
Introduce, thus, the multipole moments
\begin{align}
 \epsilon_{\ell,m}
 &\equiv
 \int d^2x\,
 r^{\ell}e^{im\varphi}\,
 \delta e(\bm x),
 \label{app:moment-definition}
\end{align}
so that keeping the linear response and only its
lowest radial mode gives
\begin{align}
 F_{n,{\rm ev}}
 &\simeq
 g_{n0}\epsilon_{n,n}.
 \label{app:leading-long-response}
\end{align}
This is the precise content of the leading eccentricity-like scaling used
in the main text.  The first corrections are of two different types.
Higher radial modes enter already within linear response,
\begin{equation}
 F_n^{\rm linear}
 =
 g_{n0}\epsilon_{n,n}
 +g_{n1}\epsilon_{n+2,n}
 +\cdots,
 \label{app:higher-radial-response}
\end{equation}
while nonlinear response produces products of modes whose harmonics add to
$n$.  Consequently, the mean-squared harmonic
$\langle f_n^2\rangle_{\rm ev}=\langle|F_n|^2\rangle_{\rm ev}$ begins as
\begin{align}
 \left\langle|F_{n, \rm ev}|^2\right\rangle_{\rm ev}
 ={}&
 |g_{n0}|^2
 \left\langle|\epsilon_{n,n}|^2\right\rangle_{\rm ev}
 \nonumber\\
 &\hspace{-1 cm}+
 2\operatorname{Re}\!\left[
 g_{n0}g_{n1}^*
 \left\langle \epsilon_{n,n}\epsilon_{n+2,n}^*\right\rangle_{\rm ev}
 \right]
 +\cdots
 +\mathcal O(\delta e^3).
 \label{app:EEC-corrections}
\end{align}
The second term illustrates that the leading radial correction is
generally a cross-covariance between the lowest and next radial moments,
rather than only the square of the next moment. This is the term schematically denoted by $ \mathcal O (r^{2n+2}(\delta e)^2)$ in the main text. Instead, the terms denoted
$\mathcal O(\delta e^3)$ arise from the interference of the linear and
quadratic response.

In summary, the response construction used in the main text can be viewed
as the following sequence:
\begin{equation}
 \begin{gathered}
 \mathcal H
 \xrightarrow{\text{expand in }\delta\Phi}
 \bigl\{H_{(1)},H_{(2)},\ldots\bigr\},
 \\
 \bigl\{H_{(p)}\bigr\}
 \xrightarrow{\mathcal P_n\circ\mathcal L}
 \bigl\{h_n^{(1)},h_n^{(2)},\ldots\bigr\},
 \\
 \bigl\{h_n^{(p)}\bigr\}
 \xrightarrow[\text{linear response ($p=1$)}]{\text{long wavelengths}}
 \bigl\{g_{n0},g_{n1},\ldots\bigr\}.
 \end{gathered}
 \label{app:ordered-response-summary}
\end{equation}
The leading result retained in the main text keeps $H_{(1)}$ in the
fluctuation expansion and $g_{n0}$ in the radial expansion.  All
short-distance collision dynamics, including the transitions between the
different stages of the evolution, is absorbed into the response
coefficients, whereas the dependence on the long-wavelength initial
geometry is carried by the multipole moments $\epsilon_{n+2\ell,n}$.


\section{3. Rotational covariance of the response kernels}
\label{app:rotational-covariance}

Here we derive the constraint that rotational symmetry places on the
kernels mapping initial perturbations to the asymptotic energy-flow
harmonics, and in particular the angular form quoted in
Eq.~\eqref{eq:angularResponse}.  The only inputs are an azimuthally
symmetric mean initial configuration and the rotational covariance of the
complete evolution and measurement chain; no hydrodynamic approximation is
involved.  We work directly with the scalar energy density $e(\bm x)=\bar e(r)+\delta e(\bm x)$. The argument generalizes for channels with vector or tensor indices.

Let $R_\alpha$ denote an active counterclockwise rotation by an angle
$\alpha$ in the transverse plane.  On a scalar field it acts as
\begin{equation}
  [U_\alpha e](\bm x)
  \equiv e(R_{-\alpha}\bm x),
  \label{eq:app-active-scalar-rotation}
\end{equation}
moving a feature located at azimuth $\varphi_0$ to $\varphi_0+\alpha$; the
symmetric background satisfies $U_\alpha\bar e=\bar e$.  Denote by
$\mathcal F_e(\phi)$ the asymptotic azimuthal energy-flow distribution
produced by the initial configuration $e$, i.e., the composition of the
full dynamical evolution with the light-ray projection, at the
pseudorapidity considered, and define
\begin{equation}
  \mathfrak F_n[e]
  \equiv\int_0^{2\pi}d\phi\,e^{in\phi}\,\mathcal F_e(\phi),
  \label{eq:app-Fscript-definition}
\end{equation}
so that $\mathfrak F_n[e_{\rm ev}]=F_{n,{\rm ev}}$.  Rotational covariance
of the dynamics and of the asymptotic measurement means that rotating the
initial state and then evolving is the same as evolving first and rotating
the final energy flow,
\begin{equation}
  \mathcal F_{U_\alpha e}(\phi)
  =\mathcal F_{e}(\phi-\alpha).
  \label{eq:app-output-equivariance}
\end{equation}
Shifting the integration variable in Eq.~\eqref{eq:app-Fscript-definition}
then gives 
\begin{equation}
  \mathfrak F_n[U_\alpha e]
  =e^{in\alpha}\,\mathfrak F_n[e],
  \label{eq:app-Fn-character}
\end{equation}
the plus sign in the phase being fixed by the convention
$F_n=\int d\phi\,e^{in\phi}\mathcal F(\phi)$.

Now expand about the symmetric background, as in the main text,
\begin{equation}
  \mathfrak F_n[\bar e+\delta e]
  =\sum_{p\geq0}\frac{1}{p!}
  \int\prod_{a=1}^{p}d^2x_a\,
  h^{(p)}_{n}(\bm x_1,\ldots,\bm x_p)
  \prod_{a=1}^{p}\delta e(\bm x_a),
  \label{eq:app-kernel-taylor}
\end{equation}
with kernels symmetric under permutations of their arguments.  Because
$U_\alpha(\bar e+\delta e)=\bar e+U_\alpha\delta e$, applying
Eq.~\eqref{eq:app-Fn-character} to this expansion, changing variables to
$\bm x_a=R_\alpha\bm y_a$, and matching order by order in
the arbitrary perturbation $\delta e$ yields the identity
\begin{equation}
  h^{(p)}_{n}(R_\alpha\bm x_1,\ldots,R_\alpha\bm x_p)
  =e^{in\alpha}\,
  h^{(p)}_{n}(\bm x_1,\ldots,\bm x_p).
  \label{eq:app-kernel-rotation-derived}
\end{equation}

For $p=1$, writing $\bm x=(r,\varphi)$, this immediately gives
\begin{equation}
  h_n^{(1)}(r,\varphi)=e^{in\varphi}\rho_n(r)\,
  \qquad
  \rho_n(r)\equiv h_n^{(1)}(r,0),
  \label{eq:app-linear-kernel-solution}
\end{equation}
which is the desired result. Rotational symmetry fixes the
entire angular dependence of the linear kernel, and all of the dynamics
 resides in the radial function $\rho_n(r)$.

\section{4. Thickness weighting and nuclear multipole}
\label{app:thickness-weighted-multipoles}

Here we reduce the covariance of the deposited energy density to the
nuclear multipole correlator of the main text.  We consider a central
collision of two statistically independent identical nuclei, with thickness
functions
\begin{equation}
 t^X_{\rm ev}(\bm x)
 =\bar t(r)+\delta t^X_{\rm ev}(\bm x),
 \qquad X=+,-,
 \label{eq:app-tbar-split}
\end{equation}
where $r=|\bm x|$, the common mean profile is azimuthally symmetric, and
$\langle\delta t^X_{\rm ev}(\bm x)\rangle_{\rm ev}=0$.  Normalization
constants are kept explicit where instructive.  Inserting
Eq.~\eqref{eq:app-tbar-split} into the local deposition map,
$e_{\rm ev}(\bm x)=\mathcal C_0^2\,t^+_{\rm ev}(\bm x)\,t^-_{\rm ev}(\bm x)$,
gives
\begin{align}
 \delta e_{\rm ev}(\bm x)
 =\mathcal C_0^2\bigl [
 \bar t(r)\,[\delta t^+_{\rm ev}(\bm x)
              +\delta t^-_{\rm ev}(\bm x)] +\delta t^+_{\rm ev}(\bm x)\,
       \delta t^-_{\rm ev}(\bm x)\bigr ].
 \label{eq:app-delta-e-exact}
\end{align}
Since the mean of the last term vanishes because the two incoming states
are independent, upon averaging one obtains
\begin{align}
 &\left\langle
 \delta e_{\rm ev}(\bm x_1)\,
 \delta e_{\rm ev}(\bm x_2)
 \right\rangle_{\rm ev}
=\mathcal C_0^4\bigl[
 2\,\bar t(r_1)\bar t(r_2)\,C_t(\bm x_1,\bm x_2)
 + \ldots \bigr],
 \label{eq:app-product-covariance-exact}
\end{align}
where
$C_t(\bm x_1,\bm x_2)\equiv\langle\delta t^X_{\rm ev}(\bm x_1)\,
\delta t^X_{\rm ev}(\bm x_2)\rangle_{\rm ev}$
is the common thickness covariance of either nucleus. The omitted term is of fourth order in the
fluctuation amplitude and is dropped.  The
energy-density covariance thus carries one factor of the mean profile at
each of its two points.

In terms of nucleonic degrees of freedom, the thickness is
the one-body density operator
$\widehat t(\bm x)=\sum_{i=1}^{A}\delta(\bm x-\widehat{\bm x}_i)$,
with transverse coordinates relative to the nuclear center of mass, and
the event ensemble of thickness fluctuations is the measurement ensemble
of the incoming nuclear state (omitting hat symbols),
\begin{equation}
 \bar t=\bigl\langle t\,\bigr\rangle_{\Psi},
 \qquad
 C_t(\bm x_1,\bm x_2)
 =\bigl\langle
 \delta t(\bm x_1)\,
 \delta  t(\bm x_2)
 \bigr\rangle_{\Psi},
 \label{eq:app-quantum-thickness-covariance}
\end{equation}
with $\delta  t\equiv  t-\bar t$.  Following the steps of the previous sections, it is then straightforward to arrive at
\begin{equation}
 \left\langle f_n^2\right\rangle_{\rm ev}
 \simeq 2
|\mathcal C_0^2\,g_{n0}|^2
 \bigl\langle
 \widehat{\mathcal Q}^{[\bar t]}_{n}\,
 \widehat{\mathcal Q}^{[\bar t]\dagger}_{n}
 \bigr\rangle_{\Psi},
 \label{eq:app-weighted-EEC}
\end{equation}
where we introduce the
\emph{thickness-weighted} multipole,
\begin{equation}
 \widehat{\mathcal Q}^{[\bar t]}_{n}
 \equiv
 \int d^2x\,
 \bar t(r)\,r^ne^{in\varphi}\,  t(\bm x)
 =\sum_{i=1}^{A}
 \bar t( r_i)\,
  r_i^{\,n}e^{in\varphi_i}.
 \label{eq:app-weighted-Q-operator}
\end{equation}
  Writing out the double sum,
\begin{align}
 \bigl\langle
 \widehat{\mathcal Q}^{[\bar t]}_{n}
 \widehat{\mathcal Q}^{[\bar t]\dagger}_{n}
 \bigr\rangle_{\Psi}
 ={}&\Bigl\langle\sum_{i}
 \bar t( r_i)^2\, r_i^{\,2n}\Bigr\rangle_{\Psi}
 \nonumber\\
 &+\Bigl\langle\sum_{i\neq j}
 \bar t(  r_i)\bar t( r_j)\,
   r_i^{\,n}  r_j^{\,n}
 e^{in( \varphi_i- \varphi_j)}\Bigr\rangle_{\Psi},
 \label{eq:app-one-two-body-split}
\end{align}
 the genuine two-body component of the nuclear correlator enters
weighted by $\bar t(r_1)\bar t(r_2)$, while the $i=j$ term is a one-body
average.

What is the effect of these weights?  A smooth azimuthally symmetric profile
admits an expansion,
\begin{equation}
 \bar t(r)
 =b_0+b_1r^2+b_2r^4+\cdots,
 \label{eq:app-tbar-even-expansion}
\end{equation}
therefore, multiplication by the radial scalar $\bar t(r)$
 cannot change the angular harmonic. It only shifts radial
powers, $r^n\to r^{n+2j}$, which is precisely the tower of subleading
terms already discarded in $\rho_n(r)\simeq g_{n0}r^n$ due to dynamical considerations.  Consistency requires then
truncating both expansions together. As a simple illustration, a uniform liquid drop of radius $R$ and density
$\rho_0$ has projected thickness
\begin{align}
 \bar t(r)
 &=2\rho_0\sqrt{R^2-r^2}\;\Theta(R-r)
 \nonumber\\
 &=2\rho_0R\left[
 1-\frac{r^2}{2R^2}-\frac{r^4}{8R^4}
 -\frac{r^6}{16R^6}-\cdots
 \right]
 \label{eq:app-liquid-drop-expansion}
\end{align}
for $r<R$.  Its leading departure from a constant is an inverted parabola
with a negative coefficient, $b_1=-b_0/(2R^2)$, reflecting a weight that is
maximal in the center and falls off toward the surface.

In leading radial order, therefore, $\bar t$ is
absorbed into the matching coefficient, and
\begin{equation}
 \left\langle f_n^2\right\rangle_{\rm ev}
 \simeq
 2\,|\mathcal C_0^2\,b_0\,g_{n0}|^2
 \left\langle
 \widehat{\mathcal Q}_{n}\,
 \widehat{\mathcal Q}_{n}^{\dagger}
 \right\rangle_{\Psi},
 \qquad n\geq2,
 \label{eq:app-leading-Q-reduction}
\end{equation}
which is the anticipated result with
$\widehat{\mathcal Q}_{n}\equiv\sum_i r_i^{\,n}
e^{in \varphi_i}$ the bare maximally projected multipole operator.

\end{document}